\documentclass[aps,prd,preprint,groupedaddress,noshowkeys,noshowpacs]{revtex4}
\usepackage{subfigure}
\usepackage{epstopdf} 
\usepackage{graphicx} \usepackage{rotating}
\def\beq{\begin{equation}}
\def\eeq{\end{equation}}
\def\bea{\begin{eqnarray}}
\def\eea{\end{eqnarray}}

\parskip=\itemsep

 \def\pom{{I\!\!P}} \vskip1cm 
\usepackage{color} \usepackage{pdflscape} 
\definecolor{blue}{rgb}{0.3,0.3,0.9} 
\begin{document} 
\begin{center} 
{\bf \large DIFFRACTION AS A CRITICAL INGREDIENT IN SOFT SCATTERING \\ 
\vskip0.2cm Uri Maor}\\ 
\vskip 0.2cm 
School of Physics and Astronomy, Tel Aviv University, ISRAEL\\ 
\end{center} 
\vskip0.3cm 
{\bf Abstract:} The roll of diffraction in the formulation of soft scattering 
is investigated aiming to assess the resent SD TOTEM results. \\
\vskip0.3cm 
\section {INTRODUCTION} 
s channel Unitarity screening considerations date back to 
the ISR epoch. 40 years latter, theoretical estimates of soft 
scatterings channels at the TeV-scale require a unified analysis of 
elastic and diffractive scatterings (Good-Walker (GW)\cite{GW})
coupled to s and t unitarity screenings. In the following  
I shall utilize the GLM model\cite{GLM}.\\ 
The lowest order of 
s-channel unitarity bound on $a_{el}(s,b)$ 
is obtained from a diagonal re-scattering matrix,\\ 
$2Im a_{el}(s,b)={\mid}a_{el}(s,b){\mid}^2+G^{in}(s,b)$. Its 
general solution is: $a_{el}(s,b)=i\left(1-e^{-\Omega(s,b)/2}\right)$\\   
and $G^{in}(s,b)=1\,- e^{-\Omega(s,b)}$. Arbitrary $\Omega$ leads to a 
bound $\mid a_{el}(s,b)\mid\leq 2$. It  
considerably over estimates TOTEM's $\sigma_{tot}$ and $\sigma_{el}$ 
preseted in this meeting. 
A much better output is provided by Glauber's eikonal approximation  
leading to a consequent $\mid a_{el}(s,b)\mid\leq 1$, identical to   
the black disc bound. 
The screened cross sections are:\\ 
$\sigma_{tot}=2\displaystyle\int d^2 b, 
\left(1-e^{-\Omega(s,b)/2}\right)$, $\sigma_{el}=\displaystyle\int d^2 
b \left(1-e^{-\Omega(s,b)/2}\right)^2$, 
$\sigma_{inel}=\displaystyle\int d^2 b 
\left(1-e^{-\Omega(s,b)}\right)$.\\ 
In a single channel model, 
the unitarity bound is initiated by the s-channel black bound 
and the $ln^2(s)$ expanding amplitude radius. The consequent 
Froissart-Martin bound\cite{Froissart} is: $\sigma_{tot} \leq C ln^2(s/s_0).\, 
s_0=1GeV^2$ and $C\,\propto\,1/2m^2_{\pi}\simeq 30mb$. 
C is far too large to be relevant.\\ 
s-channel unitarity implies: $\sigma_{el} \leq \frac{1}{2}\sigma_{tot}$ 
and $\sigma_{inel} \geq \frac{1}{2} \sigma_{tot}.$ 
At saturation, $\sigma_{el}=\sigma_{inel}=\frac{1}{2} \sigma_{tot}$.\\ 
Introducing diffraction, significantly changes the 
features of s-unitarity. However, the\\ 
saturation signatures remain valid.\\ 
\section {GOOD-WALKER DECOMPOSITION} 
Consider a p-p scattering in which we identify two orthonormal states, 
a hadron $\Psi_h$ and a diffractive state $\Psi_D$. $\Psi_D$ replaces 
the continuous diffractive Fock states. GW noted that $\Psi_h$ and 
$\Psi_D$ do not diagonalize the 2x2 interaction matrix ${\bf T}$: 
Let $\Psi_1$ and $\Psi_2$ be eigen states of $\bf T.$ 
$\Psi_h=\alpha\Psi_1+\beta\Psi_2$, $\Psi_D=-\beta\Psi_1+\alpha\Psi_2$, 
$\alpha^2+\beta^2=1$. The eigen states initiate 4 $A_{i,k}$ 
elastic GW 
amplitudes $(\psi_i+\psi_k \rightarrow \psi_i+\psi_k)$. i,k=1,2. 
For initial $p({\bar p})-p$ we have $A_{1,2}=A_{2,1}$.\\ 
I shall follow the GLM definition, 
in which the mass distribution of $\Psi_D$ is not 
defined and requires an independent specification.\\ 
The elastic, SD and DD 
amplitudes in a 2 channel GW model are:\\ 
$a_{el}(s,b)=i\{\alpha^4A_{1,1}+2\alpha^2\beta^2A_{1,2}+\beta^4A_{2,2}\}$, 
$a_{sd}(s,b)=i\alpha\beta\{-\alpha^2A_{1,1}+ 
(\alpha^2-\beta^2)A_{1,2}+\beta^2A_{2,2}\}$,\, 
$a_{dd}(s,b)=i\alpha^2\beta^2\{A_{1,1}-2A_{1,2}+A_{2,2}\}$, in which,  
$A_{i,k}(s,b)=\left(1-e^{\frac{1}{2}\Omega_{i,k}(s,b)}\right)\leq 1$.\\ 
GW diffraction has distinct features:\\ 
1) The Pumplin\cite{Pumplin} bound: 
$\sigma_{el}+\sigma_{diff}^{GW} \leq \frac{1}{2}\sigma_{tot}$.\\ 
$\sigma_{diff}^{GW}$ is the sum of the GW soft diffractive cross 
sections.\\ 
2) Below saturation, 
$\sigma_{el} \leq \frac{1}{2}\sigma_{tot}-\sigma_{diff}^{GW}$ 
and $\sigma_{inel} \geq \frac{1}{2} \sigma_{tot}+\sigma_{diff}^{GW}$.\\ 
3) $a_{el}(s,b)=1,$ when and 
only when, $A_{1,1}(s,b)=A_{1,2}(s,b)=A_{2,2}(s,b)=1$.\\ 
4) When $a_{el}(s,b)=1$, 
all diffractive amplitudes at the same (s,b) vanish.\\ 
5) GW saturation signatures are valid also in the non GW sector.\\ 
6) The saturation signature, 
$\sigma_{el}=\sigma_{inel}=\frac{1}{2}\sigma_{tot}$, in a multi channel 
calculation, is coupled to $\sigma_{diff}=0$. Consequently, prior to 
saturation the diffractive cross sections stop growing and start to 
decrease with energy. This may serve as a signature of approching saturation.\\ 
\section{CROSSED CHANNEL UNITARITY} 
Translating the concepts 
presented into a viable phenomenology requires a specification of 
$\Omega(s,b)$, for which Regge Pomeron ($\pom$) 
theory is a powerful tool. 
Mueller\cite{Mueller} applied 3 body unitarity to equate the cross section of  
$a\,+\,b \rightarrow M_{sd}^2\,+\,b$ to the 
triple Regge diagram $a\,+b\,+\bar{b} \rightarrow 
a\,+\,b\,+\,\bar{b},$ with a leading 3$\pom$ vertex term. 
The 3$\pom$ approximation is valid when 
$\frac{m_p^2}{M_{sd}^2}\,<<\,1$ and $\frac{M_{sd}}{s}\,<<\,1$.  
The leading energy and mass dependences are 
$\frac{d\sigma^{3\pom}}{dt\,dM_{sd}^2} \propto s^{2\Delta_{\pom}} 
(\frac{1}{M_{sd}^2})^{1+\Delta_{\pom}}$.\\ 
Mueller's 3$\pom$ approximation for non GW diffraction 
is the lowest order of t-channel 
multi $\pom$ interactions, compatible with t-channel unitarity.\\  
t-channel screening results in a distinction between 
GW and non GW  diffraction. 
Recall that, unitarity screening of GW ("low mass") 
diffraction is controled by eikonalization, while the 
screening of non GW ("high mass") diffraction is controled by the 
survival probability. 
Note that the relationship between GW  
and Mueller's diffraction modes needs further study.\\
\section{THE PARTONIC POMERON} 
Current $\pom$ models differ in details, but have in 
common a relatively large adjusted input $\Delta_{\pom}$ and a 
diminishing $\alpha_{\pom}^{\prime}$. 
Traditionally, $\Delta_{\pom}$ determines the energy dependence of the 
total, elastic and diffractive cross sections while 
$\alpha^{\prime}_{\pom}$ determines the forward slopes. This picture is 
modified in updated $\pom$ models in which s and t unitarity 
screenings induce a much smaller $\pom$ intercept at t=0, denoted 
$\Delta^{eff}_{\pom}$, which gets smaller with energy.
The exceedingly small fitted $\alpha_{\pom}^{\prime}$ 
implies a partonic description of the $\pom$ which leads to a pQCD 
interpretation.\\ 
Gribov's partonic Regge theory\cite{VG} provides the 
microscopic sub structure of the $\pom$ where the slope of the $\pom$ 
trajectory is related to the mean transverse momentum of the\\ 
partonic dipoles constructing the Pomeron.  
$\alpha_{\pom}^{\prime}\,\propto\,1/< p_t >^2$.\\ 
Accordingly: $\alpha_S \propto \pi/ln \left(< p_t^2 >/\Lambda_{QCD}^2 \right) << 1.$\\ 
We obtain a $\pom$ with hardness changing continuesly from hard (BFKL like) 
to soft (Regge like). This is a non trivial relation as the soft $\pom$ 
is a simple moving pole in J-plane, while  
the BFKL hard $\pom$ is a branch cut, 
approximated as a simple pole with 
$\Delta_{\pom}=0.2-0.3$\\ 
and $\alpha_{\pom}^{\prime}\simeq 0$.\\ 
GLM\cite{GLM} and KMR\cite{KMR} models are rooted in Gribov's partonic $\pom$ 
theory with a hard pQCD $\pom$ input. It is softened by unitarity 
screening (GLM), or the decrease of its partons' transverse momentua 
(KMR). The two definitions are correlated. GLM and KMR have a bound 
of validity at 60-100 TeV implied by their approximations.\\ 
\section{UNITARITY SATURATION} 
Unitarity saturation is coupled to 3 experimental signatures:\\ 
$\frac{\sigma_{inel}}{\sigma_{tot}}$ = 
$\frac{\sigma_{el}}{\sigma_{tot}} = 0.5,$ \,\, 
$\frac{\sigma_{tot}}{B_{el}} = 9\pi,$ \,\,  
$\sigma_{diff}=0.$\\ 
Following is p-p TeV-scale data 
relevant to the assessment of saturation:\\ 
{\bf CDF(1.8 TeV):}
$\sigma_{tot}=80.03 \pm 2.24 mb,\,\, \sigma_{el}=19.70 \pm 0.85 mb,\,\, 
B_{el}=16.98 \pm 0.25 GeV^{-2}.$\\ 
{\bf TOTEM(7 TeV):} 
$\sigma_{tot}=98.3 \pm 0.2(stat) \pm 2.8(sys) mb,$ \,\, 
$\sigma_{el}=24.8 \pm 0.2(stat) \pm 2.8(sys) mb,$ \,\, $B_{el}=20.1 \pm 
0.2(stat) \pm 0.3(sys) GeV^{-2}.$\\ 
{\bf AUGER(57 TeV):} 
$\sigma_{tot}=133 \pm 13(stat) {\pm}^{17}_{20}{sys} \pm 16(Glauber) mb,$\\ 
$\sigma_{inel}=92 \pm 7(stat) \pm^{9}_{11} (sys) \pm 16(Glauber)mb.$\\ 
Note that AUGER output margin of error is large!\\
$\sigma_{inel}/\sigma_{tot}$=0.754(CDF), 0.748(TOTEM), 0.692(AUGER).\\ 
The numbers suggest a very slow approach toward 
saturation, well above the TeV-scale.\\ 
Since multi-channel models are bounded to the TeV-scale, I am limited  
to single\\ 
channel models above 100 TeV. To this end I quote a calculation 
by Block and Halzen\cite{BH}, 
who have checked the predictions of their model at 
the Planck-scale (1.22$\cdot10^{16}$TeV). They obtain 
$\sigma_{inel}/\sigma_{tot}=1131mb/2067mb$ = 0.547.\\ 
It re-enforces the conclusion that saturation will be attained, 
if at all, at non realistic energies.\\ 
\section{TOTEM RECENT SD DATA}      
As noted, the predicted vanishing of the diffractive cross sections 
at saturation implies that $\sigma_{sd}$, which up to TOTEM 
grows slowly with energy, will eventually start to reduce.\\ 
This may serve as an early signature that saturation is being 
approached.\\  
The preliminary TOTEM measurement (reported in this meeting) of 
$\sigma_{sd}=6.5\pm 1.3 mb$, corresponding to\\ 
$3.4 < M_{sd} < 1100 GeV$ and $2.4 \cdot 10^{-7} < \xi < 0.025$, suggests a 
radical change in the energy dependence of $\sigma_{sd}$, which is 
smaller than its value at CDF.\\
This feature, if correct, suggests a much faster 
approach toward unitarity saturation than suggested by 
$\frac{\sigma_{inel}}{\sigma_{tot}}.$\\ 
TOTEM diffractive data is very preliminary. Regardless, the compatibility 
between the information derived from different channels of soft 
scattering deserves a very careful study! \\
{\bf Acknowledgement}\\
This work is partially supported by UP8 of the hadron physics 
of the 8th EU program period.\\

\end{document}